\documentclass{v17dm}

\usepackage[T1]{fontenc}
\usepackage{xspace}

\def\be{\begin{equation}}
\def\ee{\end{equation}}
\def\bea{\begin{eqnarray}}
\def\eea{\end{eqnarray}}

\newcommand{\Photo}{}

\begin{document}
\vspace*{4cm}
\title{Neutrino telescope searches for dark matter in the Sun}

\author{Pat Scott}

\address{Fundamental Physics Section, Department of Physics, Imperial College London, Blackett Laboratory, Prince Consort Road, London SW7 2AZ, UK}

\maketitle\abstracts{
I give a brief review of a few recent developments and future directions in the search for dark matter using high-energy neutrinos from the Sun.  This includes the ability to recast neutrino telescope limits on nuclear scattering of dark matter to arbitrary new theories, and new calculations of the solar atmospheric background relevant to such searches. I also touch on applications to global searches for new physics, and prospects for improving searches for asymmetric dark matter in the Sun.
}

\section{Current status}

High-energy neutrinos from the Sun provide one of the cleanest potential discovery channels for weakly-interacting dark matter (DM).  Weakly-interacting DM particles passing through the Sun are expected to scatter on solar nuclei.  Some of these collisions reduce the kinetic energy of the DM particle enough for it to become gravitationally bound to the Sun, causing it to return on a bound orbit and undergo subsequent scattering, eventually thermalising and settling down to the solar core.  If DM is able to annihilate, either with itself of with anti-DM captured in a similar manner, high-energy SM particles will be produced in the solar core.  Even if neutrinos are not amongst those particles produced in the annihilation hard process, they will still be generated with quite high energies in the decay and subsequent interaction of other SM particles with nuclei in the Sun.  Unlike the other SM particles, these GeV-scale neutrinos are then able to travel unhindered from the centre of the Sun to the surface, and across space to Earth, where they may be detected with terrestrial experiments.

The directionality of the signal is the primary means by which it can be distinguished from the atmospheric neutrino background, caused by cosmic ray interactions with the Earth's atmosphere.  The only known background to the signal is therefore the analogous production of high-energy neutrinos in the atmosphere of the Sun, due to interactions of cosmic rays with solar nuclei.

The capture of dark matter by the Sun typically becomes the rate-limiting step in the production of any signal, rather than the annihilation.  Searches for high-energy neutrinos from the Sun are therefore most useful for constraining the interaction cross-section of dark matter with nuclei.  Spin-dependent interactions are particularly relevant, as the Sun consists mostly of hydrogen, which possesses nuclear spin.

\begin{figure}
\includegraphics[width=0.5\textwidth]{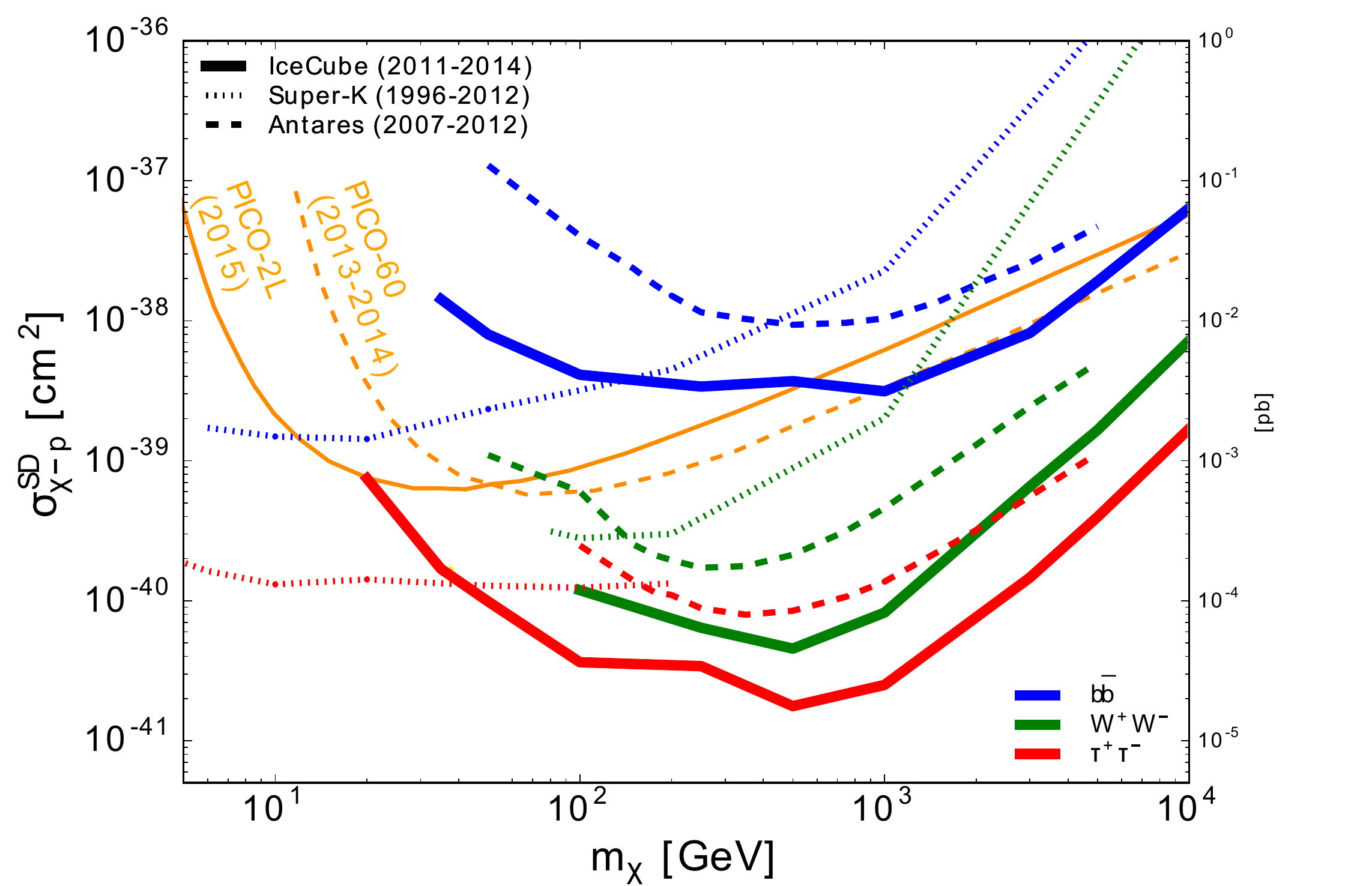}
\includegraphics[width=0.5\textwidth]{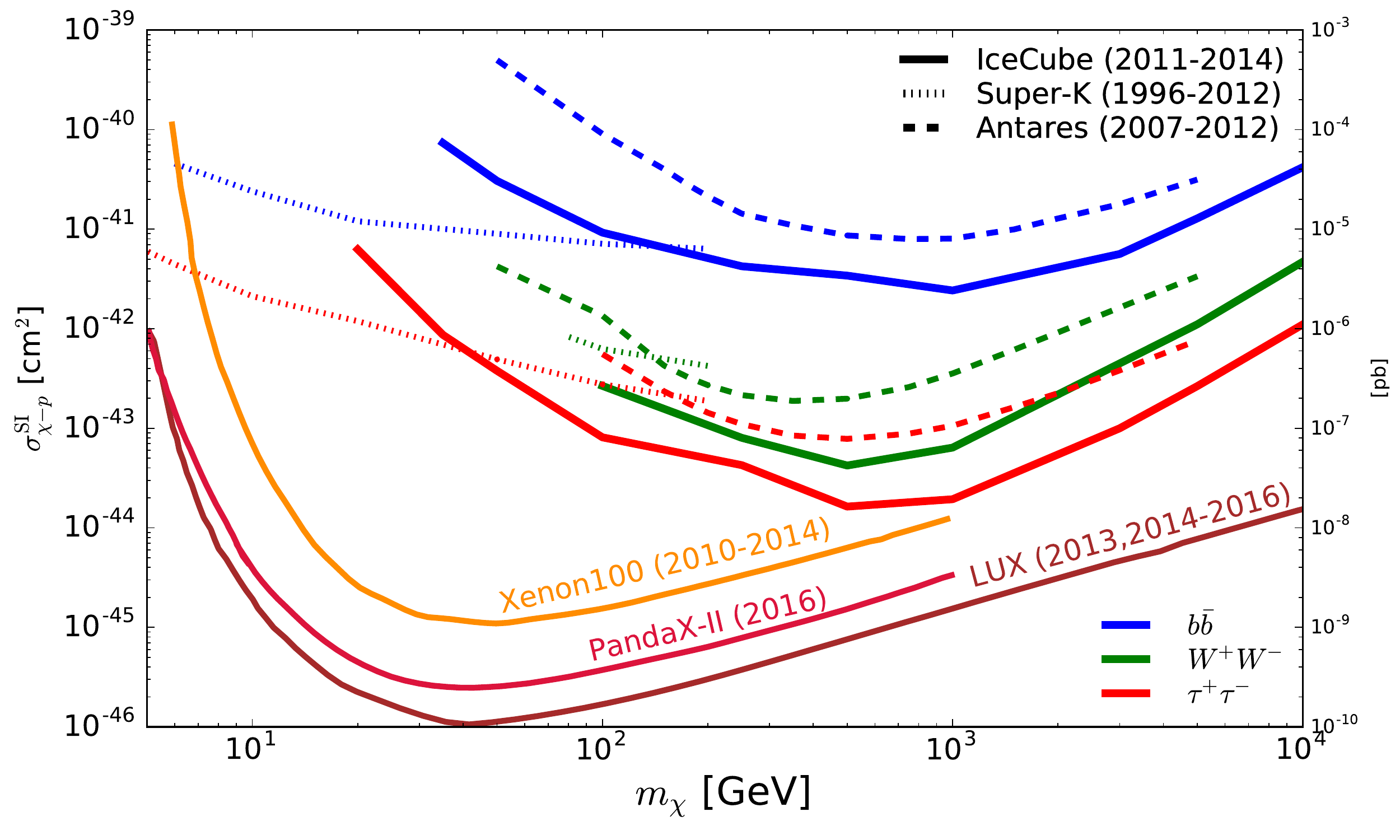}\\
\includegraphics[width=0.48\columnwidth]{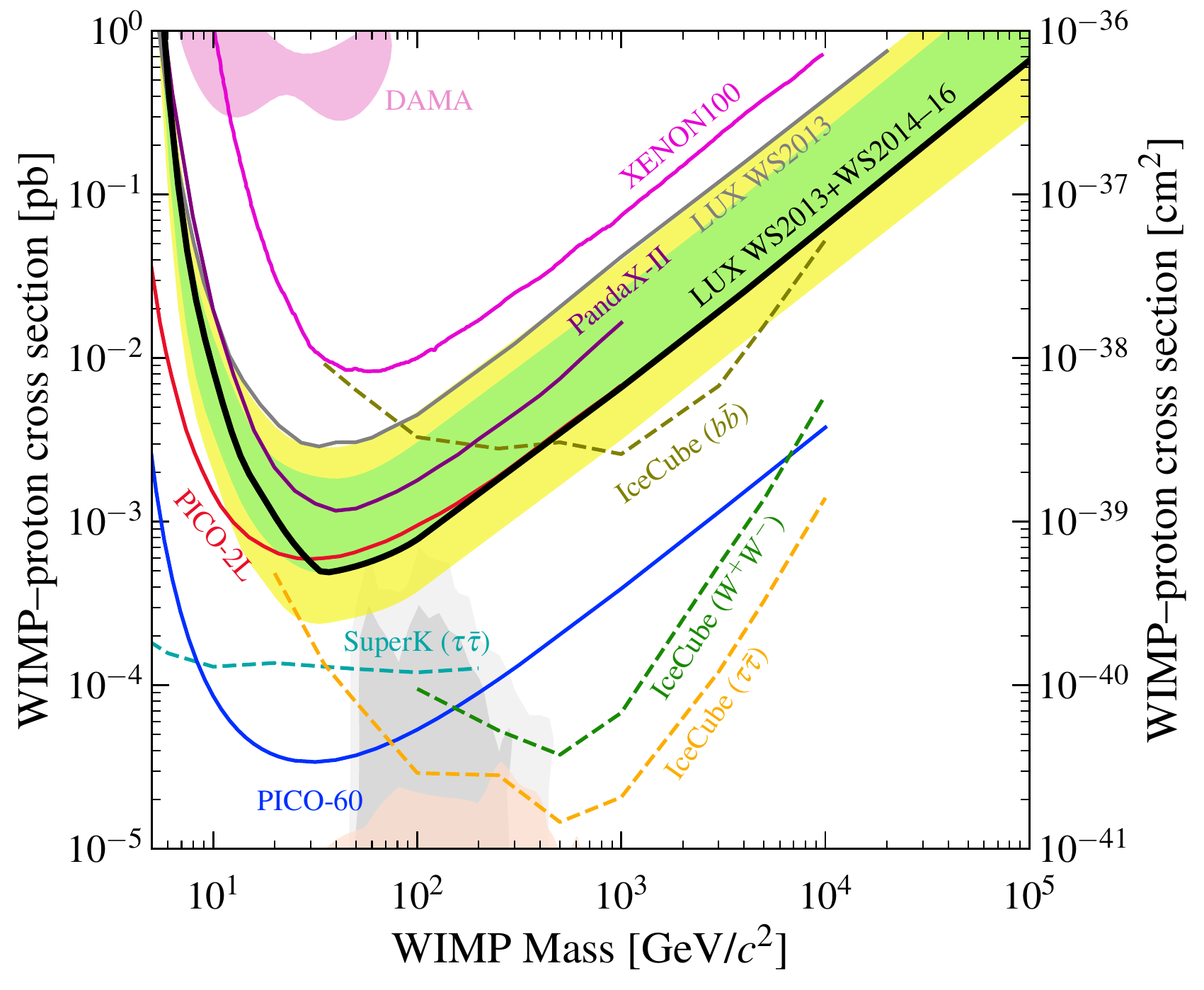}\hspace{0.03\columnwidth}
\includegraphics[width=0.48\columnwidth]{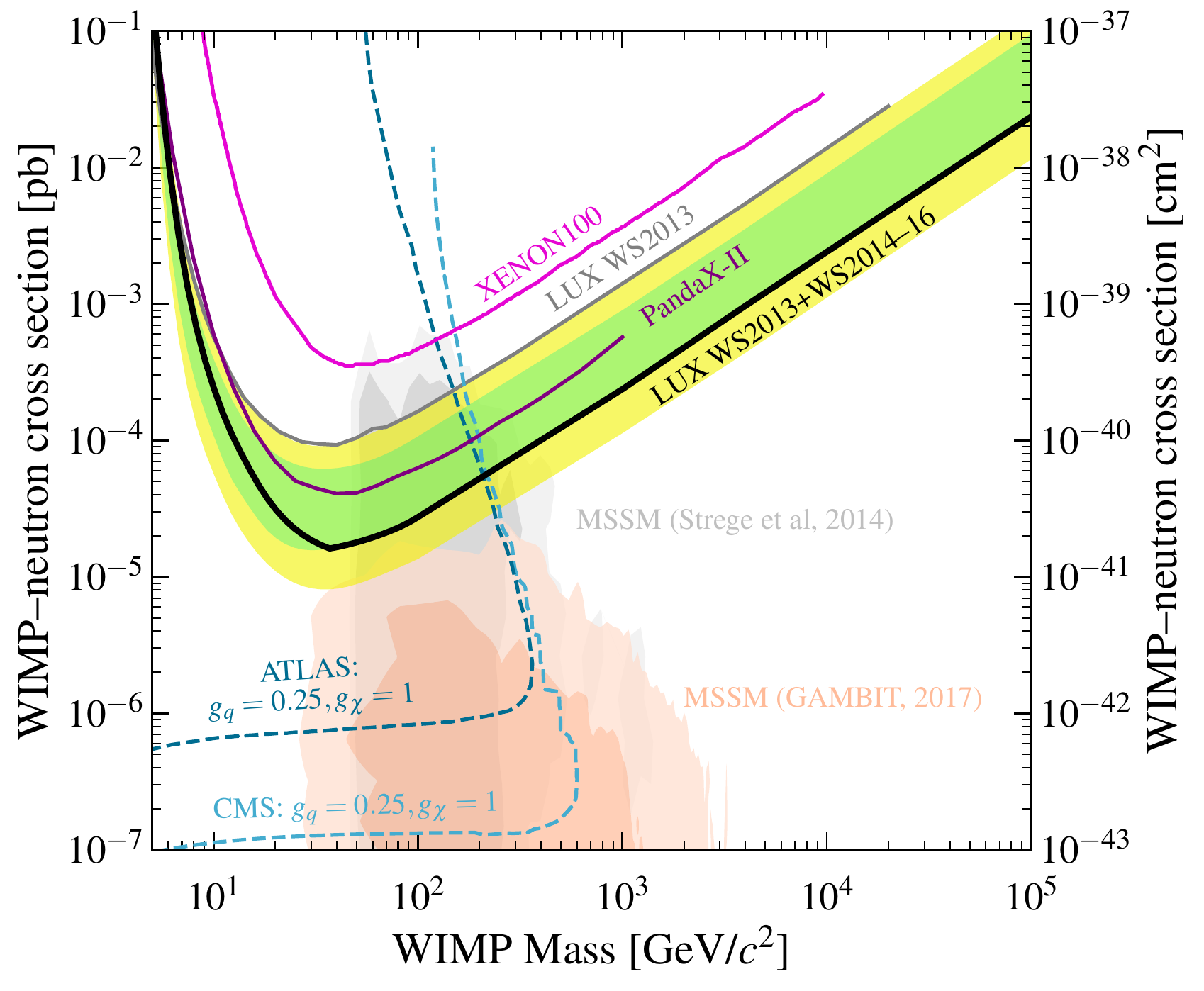}
\caption[]{Current limits on nuclear scattering of dark matter from neutrino telescopes and direct detection.  The top row shows limits from the leading neutrino telescopes on both the spin-dependent scattering cross-section with protons, and the spin-independent cross-section (with any nucleon).   The lower row compares limits on the spin-dependent interactions with protons and neutrons, illustrating the role of various direct detection and collider experiments, as well as the evolution of corresponding supersymmetric theory predictions over time.   Figures from IceCube$^{\,}$\cite{IC86} (top row) and LUX$^{\,}$\cite{LUXSD_2017} (bottom row).}
\label{fig1}
\end{figure}

Current limits from neutrino telescope and direct searches for dark matter scattering are shown in Fig. \ref{fig1}.  The IceCube neutrino telescope presently provides the leading sensitivity to spin-dependent scattering with protons at high DM masses$^{\,}$\cite{IC86}, whereas Super-Kamiokande$^{\,}$\cite{SuperK15} and PICO-60$^{\,}$\cite{PICO60_2017} have the leading sensitivity at low masses.  ANTARES$^{\,}$\cite{ANTARES16} also provides complementary constraints.  Direct searches lead the way for spin-independent interactions, and spin-dependent interactions with neutrons.

\section{Improved background calculations}

\begin{figure}
\centering
\includegraphics[width=0.5\textwidth]{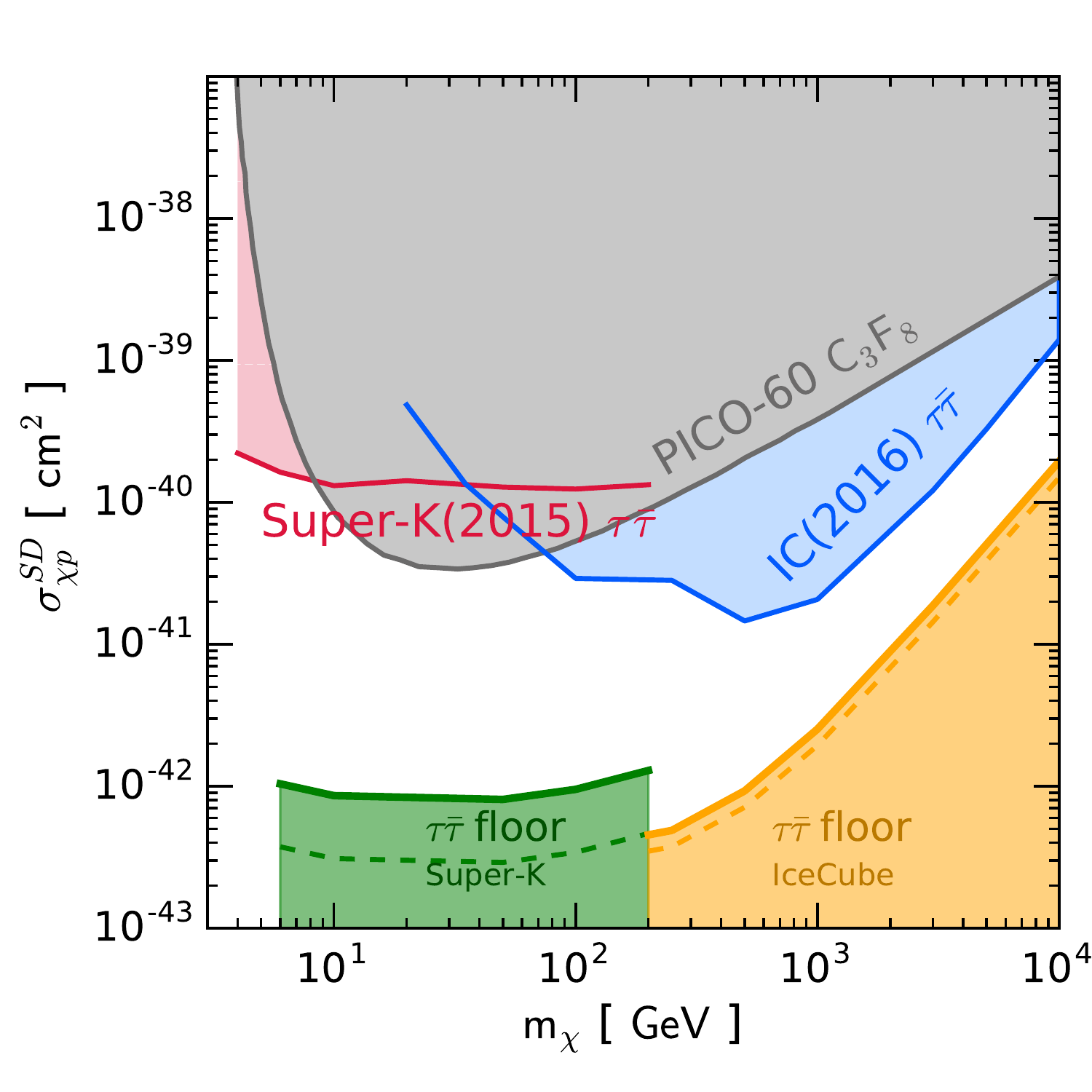}
\caption[]{Current limits on spin-dependent nuclear scattering of DM, compared to older calculations of the neutrino floor for neutrino searches toward the Sun; more accurate calculations of the floor are now available$^{\,}$\cite{Aguelles17,Edsjo17}, but give broadly similar results.  Figure from Ng et al.$^{\,}$\cite{Ng17}}
\label{fig2}
\end{figure}

Previous predictions of the background rate of high-energy neutrinos from the Sun, due to interactions of cosmic rays with nuclei in the solar atmosphere, were computed more than a decade ago.$^{\,}$\cite{Ingelman96,Hettlage00,Fogli06}  However, two more recent recalculations have appeared.$^{\,}$\cite{Aguelles17,Edsjo17}  Compared to the older predictions, the new calculations make use of modern knowledge on neutrino oscillations, production and interaction cross-sections.  One of these$^{\,}$\cite{Edsjo17} also makes use of up-to-date models of the solar composition and structure, and carries out extensive Monte Carlo simulations of neutrino production, interaction and oscillation.  Both studies (and another at the same time, based on the old flux estimates$^{\,}$\cite{Ng17}) show that the solar atmospheric background lies barely an order of magnitude below current sensitivity limits for some models (Fig.\ \ref{fig2}).  This suggests that future neutrino telescopes might be able to directly measure this irreducible `neutrino floor', and that the improved calculations of the background rates should be included in future phenomenological studies of DM scattering and annihilation in the Sun.

\begin{figure}
\centering
\includegraphics[width=0.7\textwidth]{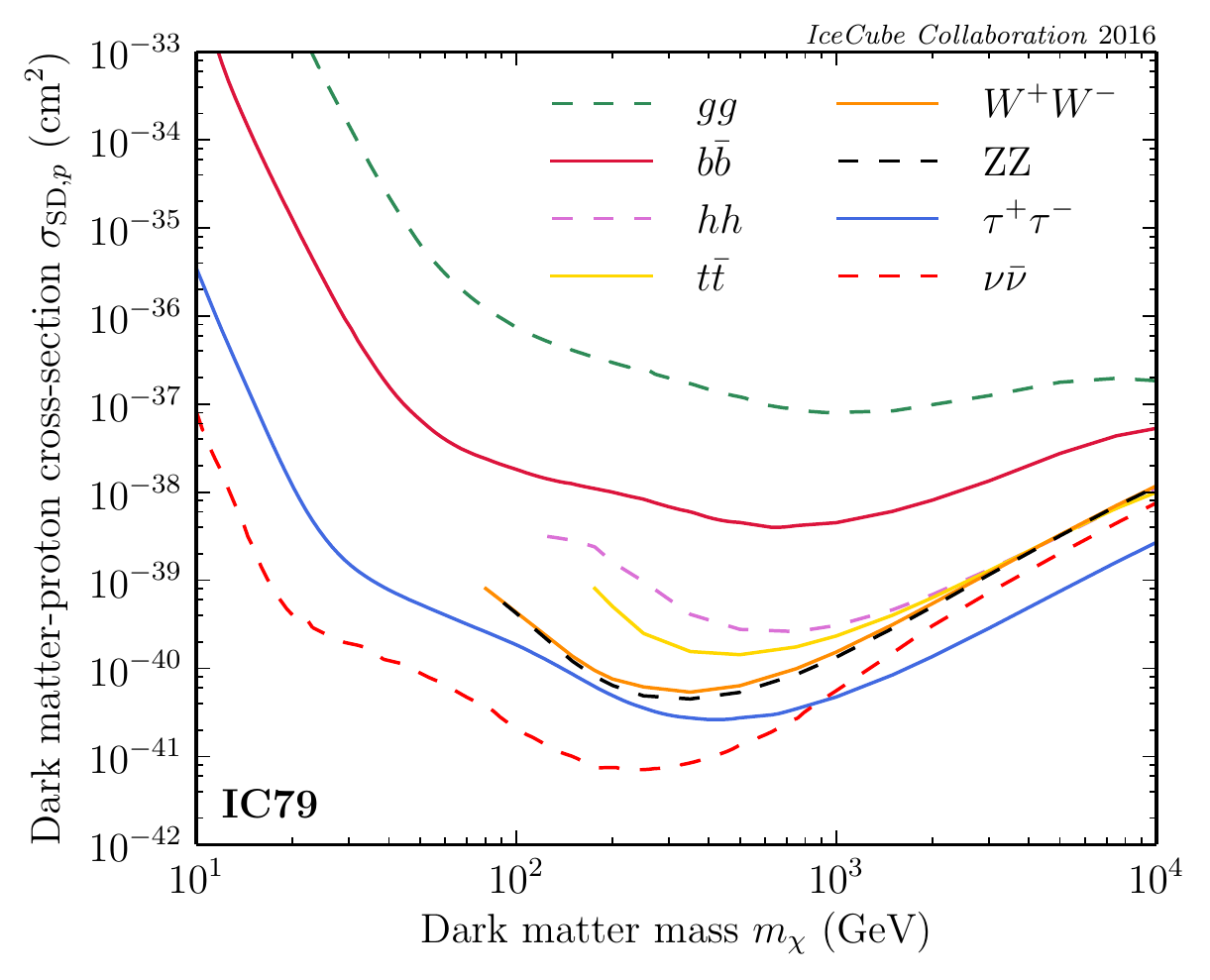}
\caption[]{90\%CL upper limits on the DM-proton nuclear scattering cross-section derived by recasting the IceCube 79-string search for DM in the Sun.  Different curves assume different annihilation final states. Figure from IceCube.$^{\,}$\cite{IC79_SUSY}}
\label{fig3}
\end{figure}

\begin{figure}
\centering
\includegraphics[width=0.7\textwidth]{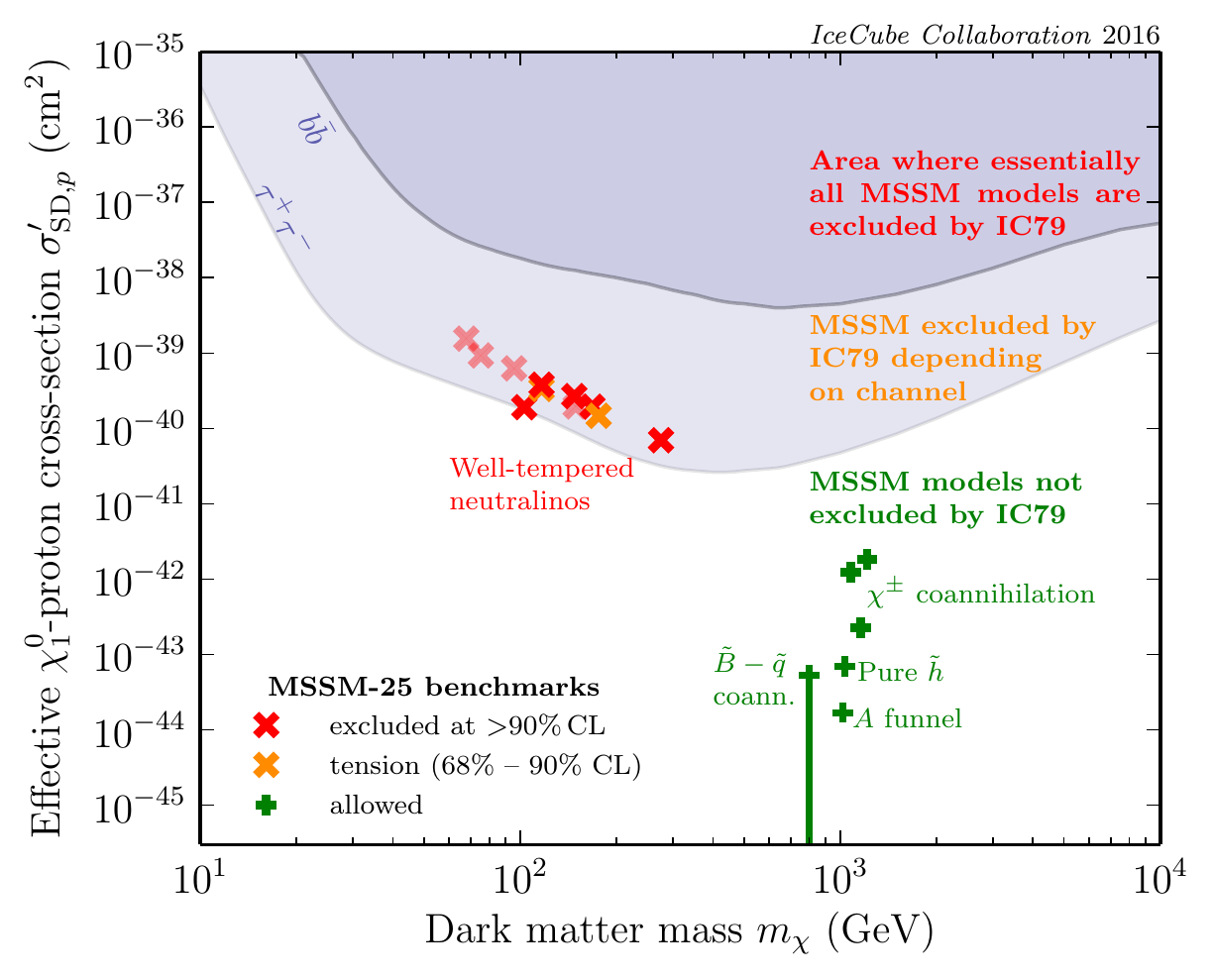}
\caption[]{25-parameter supersymmetric benchmark models from Silverwood et al.$^{\,}$\cite{Silverwood12} and Cahill-Rowley et al.$^{\,}$\cite{SM13}, colour-coded according to how strongly disfavoured they are by the recast limit from the 79-string IceCube search for DM annihilation in the Sun. The borders of the grey regions show the recast limits for typical `hard' ($\tau^+\tau^-$) and `soft' ($b\bar{b}$) spectra seen in supersymmetric models.  Figure from IceCube.$^{\,}$\cite{IC79_SUSY}}
\label{fig4}
\end{figure}

\section{Recasting efforts}

One of the major difficulties in interpreting the results of neutrino searches for DM in the Sun is the model-dependence of most published limits.  This arises from the fact that the signal prediction is highly model-dependent, as the capture rate, annihilation rate, annihilation branching fractions and resulting neutrino spectrum predicted at Earth all enter the calculation of the predicted signal in a significant way.  Traditional presentations from experiments$^{\,}$\cite{IC86,ANTARES16,IC79} give limits on scattering cross-sections as a function of DM mass, under various limiting assumptions about the dominant annihilation channel.  Recently, a more general and flexible method for presenting the results of such searches has been developed$^{\,}$\cite{IC22Methods,IC79_SUSY}, allowing existing searches to be easily recast to provide detailed and consistent constraints on alternative DM models.  This allows existing results to be converted to limits on different annihilation channels than those assumed in the original analysis (Fig. \ref{fig3}), and for them to be applied to much more complex models (Fig.\ \ref{fig4}), including arbitrary combinations of different annihilation final states and nuclear interactions (spin-dependent, spin-independent, and even more general forms).  Public software exists to perform the recast (\textsf{nulike}: \href{http://nulike.hepforge.org}{http://nulike.hepforge.org}).

This recastable form also has the advantage of providing substantial additional information compared to the traditional presentation, as it provides event-level information on neutrino/muon arrival angles and energies, and a detailed approximation to the full likelihood form of the experiment.  This is particularly important when including searches for high-energy neutrinos from the Sun in global analyses of new physics scenarios such as supersymmetry.  IceCube searches have been used in this format in the most recent reference global analyses of supersymmetric DM$^{\,}$\cite{CMSSM,MSSM} and scalar singlet DM$^{\,}$\cite{SSDM}, done in the context of the GAMBIT project.$^{\,}$\cite{gambit,ColliderBit,DarkBit,SDPBit,FlavBit,ScannerBit}

\section{Prospects for combined analysis and application to new models}

Whilst the recastable form$^{\,}$\cite{IC79_SUSY} of the 79-string IceCube data$^{\,}$\cite{IC79} is far more flexible and generally useful to the phenomenological community, there is some amount of overhead required to reduce the data to the necessary form.  For this reason, recastable versions of the 86-string IceCube search$^{\,}$\cite{IC86}, and the latest results from ANTARES$^{\,}$\cite{ANTARES16} and Super-Kamiokande$^{\,}$\cite{SuperK15} are not yet available.  It is hoped that this will soon change.  One significant driver for such a development is the prospect that the data of all three neutrino telescopes could be seamlessly combined, to give a single unified and strengthened limit.  Indeed, as soon as each of the individual datasets is available in recastable form, the combination would be extremely straightforward to perfom via the composite likelihood method.  Including the combined constraint in global analyses of searches for new physics would be similarly straightforward.

A related recasting application will be to rigorously apply neutrino telescope limits to models of asymmetric DM that exhibit both symmetric and asymmetric components$^{\,}$\cite{Murase16}, allowing strong constraints to be placed on their asymmetry parameter $r_\infty$.  Here the capture rates of such models need to be carefully determined via a low-energy effective operator treatment$^{\,}$\cite{CatenaSchwabe}, and the full range of possible operators for both scattering and annihilation, along with their interferences, taken into account.  The results of such an exercise will be especially interesting to compare to helioseismological and low-energy solar neutrino observables, given recent suggestions of a possible signal of DM from this sector.$^{\,}$$^{\,}$\cite{FrandsenSarkar,Taoso10,Vincent13,Vincent14,Vincent15,Vincent16}$^{,\,}$\footnote{See also A.\ Vincent, these proceedings.}

\section*{Acknowledgments}

I am supported by STFC (ST/K00414X/1, ST/P000762/1, ST/L00044X/1), and thank my co-authors on a number of the works discussed here.

\bibliography{DMbiblio,SUSYbiblio}

\end{document}